# The low-energy electron point source microscope as a tool for transport measurements of free-standing nanometer scale objects: application to carbon nanotubes.

P. Dorozhkin[*], H. Nejoh, D. Fujita

*National Institute for Materials Science, 1-2-1 Sengen, Tsukuba 3050047, Japan*

*Abstract*

We have developed a simple and reliable technique for two-terminal transport measurements of free-standing wire-like objects. The method is based on the low-energy electron point source microscope. The field emission tip of the microscope is used as a movable electrode to make a well-defined local electrical contact on a controlled place of a nanometer-size object. This allows transport measurements of the object to be conducted. The technique was applied to carbon nanotube ropes.

## 1. Introduction

Electrical measurements of wire-like nanometer-scale organic and inorganic objects are an important task in modern biology and applied physics. A number of experimental methods for providing local electrical contacts to such objects have been developed recently and applied for transport measurements of DNA molecules[1,2], nanotubes[3,4], metal nanowires[5] etc.

In the present paper we report the development of a new experimental technique for measuring the electrical properties of nanometer scale objects. The method is based on the low energy electron point source (LEEPS) microscope[6,2] and it extends the conventional imaging capabilities of the microscope. We realize a procedure of making a well controlled electrical contact between the field emission tip of the microscope and the nanometer-size free-standing object for performing transport measurements. The developed method has a number of advantages compared to previously applied techniques. In this work we use carbon nanotubes as objects of investigation.

## 2. Low energy electron point source microscope

The setup principle of the low-energy electron point source microscope[6,2] is presented in fig. 1. The microscope consists of a field emission tip which can be moved in X-,Y- & Z – directions, a free standing object and a plane electron detector aligned perpendicular to the tip axis. A negative voltage (~100V) is applied to the tip to produce electron field emission from the tip into vacuum. The electron beam has a conical shape and originates from a single point: the tip apex. If the free-standing object is placed into the beam, a magnified shadow image of the object is formed on the detector. The magnification of the shadow image, $k$, is defined by the ratio of tip-detector to tip-object distance: $k = D/d$ (fig. 1). By moving the tip in X- & Y – directions a required place on the sample can be chosen for imaging; moving the tip in the Z-direction allows changing magnification which can reach the value of $k \sim 10^6$.

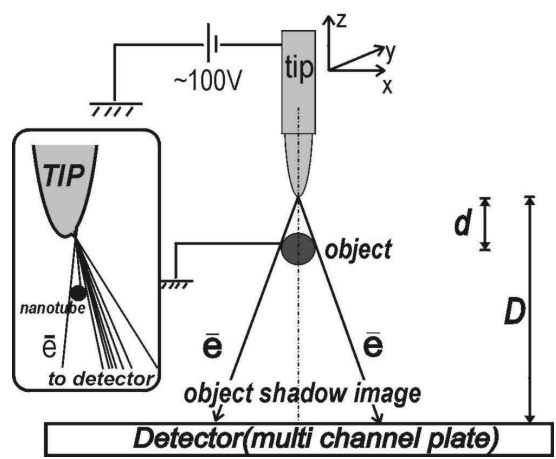

fig. 1. The setup principle of the low-energy electron point source microscope (not to scale). Inset: relative position of tip and sample at high microscope magnifications

The LEEPS-microscope has been successfully used for imaging free-standing objects such as carbon fibers and nanotubes[6,7], DNA- and RNA molecules[2,6], purple membranes[8] etc. Also electrical measurements of DNA molecules have been performed using a LEEPS-microscope equipped with an additional manipulation tip[2]. To date no experiments have been realized using the imaging field emission tip of the microscope as an electrode for contacting or manipulating nanoobjects.

Our LEEPS-microscope is based on a commercial Omicron ultra high vacuum scanning tunneling microscope[9]. X-,Y- & Z- tip movements are performed by both inertial sliders and a piezotube for coarse and fine tip positioning respectively[10]. The electron detector – a one-stage multi-channel plate (MCP) combined with a phosphor screen is placed 7 cm away from the sample as shown in fig. 1. The microscope is placed in a vacuum chamber with a working pressure of $1*10^{-9}$ Torr.

### 3. Experimental procedure & results

We used purified single-walled carbon nanotubes[11] as objects for imaging and electrical measurements. The nanotubes were deposited from a previously ultrasonicated ethanol solution onto a standard transmission electron microscope nickel microgrid, covered with a thin gold layer. The size of the grid holes was ~ 6 μm. As the nanotube ropes form a free-standing network across the holes of the microgrid, this makes them available for study by the LEEPS-microscope. Field emission tips were produced by electrochemical etching of a tungsten wire in a 2 M NaOH solution. The tips were subsequently cleaned in vacuum by $Ar^+$ sputtering and annealing at ~800°C.

Fig. 2a shows a low magnification LEEPS image of a network of carbon nanotube ropes (bundles), stretched across the holes of the microgrid (out of image view). The tip-sample distance at which the image was obtained was d ~ 5000 nm, the tip voltage was $V_{tip}$ ~ -150 V. The scale bars presented on the LEEPS-images were determined using the following procedure[6]: the tip was moved by the calibrated piezotube a known distance in X- or Y- direction and a corresponding movement of the shadow image on the screen was detected giving the precise value of the image dimensions. The diameter of observed ropes typically ranged from 2 to 30 nm which corresponds to 3 to 1000 single-walled nanotubes in one rope[12].

By moving the tip closer to the sample while keeping the field emission current constant by reducing the tip voltage, we obtained an image of a single rope (fig. 2b: d ~ 1000 nm, $V_{tip}$ ~ -90 V). This image is no longer a simple shadow image – it is accompanied by parallel alternating dark and bright lines which are interference fringes caused by diffraction of the coherent electron imaging beam by the nanotube rope[6]. Note that the rope is essentially *opaque* to electrons at the energies used (~ 100 eV)[13,14,6] and the diffraction pattern of fig. 2b can be well described by a classical Fresnel diffraction of freely propagating coherent electrons of wavelength $l$ ~ 0.1 nm ($l = h*(2meV_{tip})^{-1/2}$) by an opaque object of diameter 5 nm[6].

By carefully moving the tip further in the Z-direction towards the rope, we obtained its high magnified image (fig. 2c, $V_{tip}$ = 70 V, d ~ 250 nm). The X- & Y- tip positions were adjusted such that the image of the rope was situated on the center of the screen (marked by the white cross on fig. 2c) which lies on the tip axis - see fig. 1. This position of the image ensures us that the rope is placed right below the field emission point, even if the field emission beam originates not exactly from the top of the tip apex and has a direction different from the tip axis as shown in the inset to fig. 1.

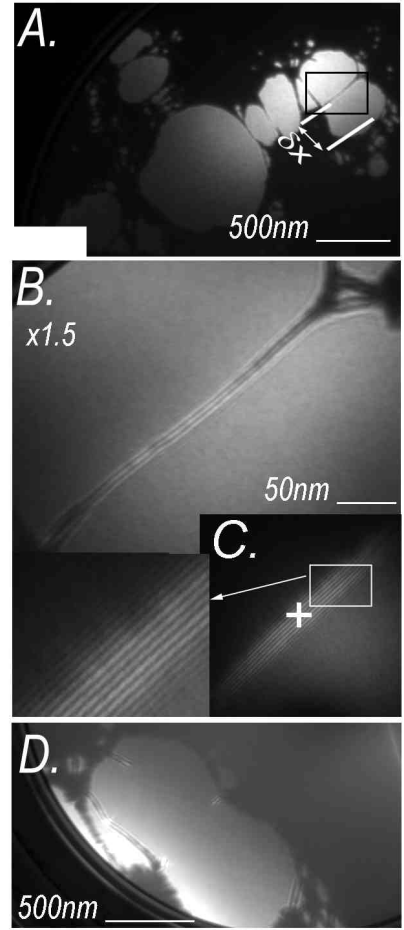

fig. 2. (a)-(c) A sequence of LEEPS-images demonstrating the process of tip approach towards a single free-standing nanotube rope with both ends fixed. The square on image (a) indicates the rope to be touched; image (b) is enlarged 1.5 times; the cross '+' on image (c) marks the geometrical center of the detector screen – the area of the rope around the center is to be contacted. After tip positioning the tip voltage was reduced to 50 mV and the tip was moved in the Z-direction towards the rope until the electrical contact was detected.

(d) Image demonstrating that the rope was destroyed after a voltage of ~5V was applied to it during transport measurements (compare to image (a) ).

Note that the highly magnified image in fig. 2c does not have any 'shadow' features any more. Indeed, the image of the rope is brighter than the background. This is a result of a strong bending of electron trajectories towards the rope, caused by a highly non-uniform distribution of electrical potential around the rope induced by the proximity of the negatively biased field emission tip[6]. In other words, the grounded rope works as an electron biprism or a 'wire lens'[15], attracting passing electrons. An analysis of electrostatic potential in the tip-sample system[16] shows that this bending of electron trajectories becomes stronger at higher magnifications, when the tip-sample distance becomes comparable to the other two geometrical parameters of the setup: the tip radius and the distance *dx* (see fig 2a) between the point of the rope imaged and the supporting media.

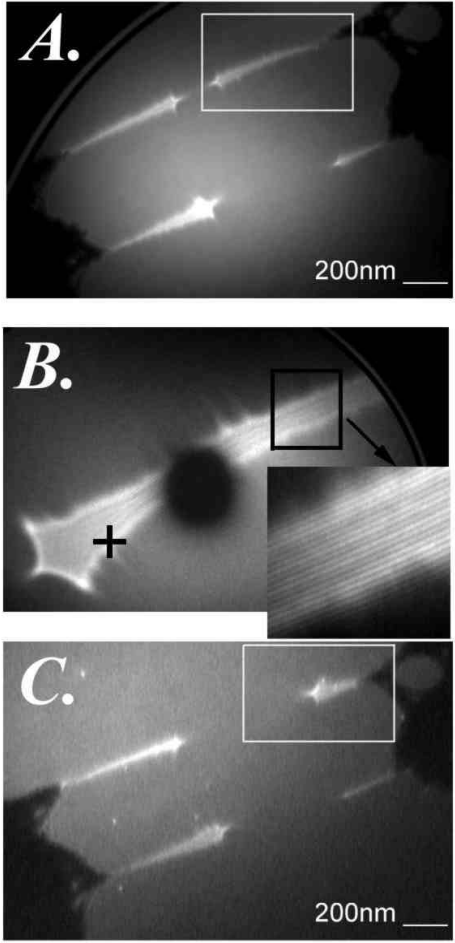

fig. 3. (a),(b) A sequence of LEEPS-images demonstrating the process of the tip approach for contacting a single free-standing nanotube rope with one free end. The cross '+' on image (b) marks the geometrical center of the detector screen; a black spot on image (b) is a defect of the detector.
(c) Image demonstrating the rope shortened after a voltage of ~6 volts was applied to it during transport measurements (compare to image (a) ).

Further movement of the tip closer to the rope, while maintaining the imaging electron beam, is hazardous since a strong electrostatic interaction between the biased tip and the grounded rope leads to instabilities of the tip-to-rope relative position and the rope can eventually break and stick to the tip. Thus, after initial tip positioning (fig. 2c) we reduced the tip voltage to a safe small value of ~ -50 mV and moved the tip in Z-direction towards the rope until electrical contact was detected by the appearance of a current between the biased tip and the ground passing through the rope. As a result of this procedure the tip touched the rope at the area centered on the screen in fig. 2c. After contacting the rope its transport characteristics could be measured. When we increased the tip voltage to a value of approximately 5 V, the electrical contact to the rope was lost. The resulting LEEPS-image taken after the loss of the contact is presented in fig. 2d, showing that the rope had been destroyed.

The same tip approach procedure as described above can also be applied in making an electrical contact to a rope with only one end fixed (fig. 3a). In this case the free end of the rope can be easily attracted towards the biased tip; thus, approaching the tip while imaging the sample requires extra precaution. Sharp field emission tips are preferable in this case, since their working voltage for field emission is lower and the corresponding tip-sample electrostatic interaction is weaker. Figures 3a,b show the process of tip approach to a rope with one free end. After touching the end of the rope with the tip, I-V characteristic of the rope was measured (fig 4, solid line). Applying a voltage of 6 V to the tip resulted in the rope being cut (fig. 3c).

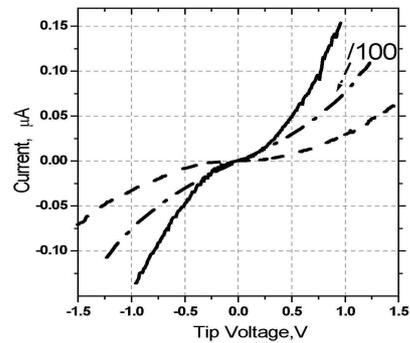

fig. 4. Solid line: I-V curve for the nanotube rope, presented in fig. 3; dashed & dash-dotted lines: typical I-V curves for high- and low-resistance nanotube ropes correspondingly; the dash-dotted curve is divided by the factor of 100.

Using the demonstrated technique we have performed electrical measurements on ~ 30 nanotube ropes with only one end fixed; the ropes had various lengths and diameters. The resistances of the nanotubes measured at low tip biases varied in a wide range from 50 k$\Omega$ to 100 M$\Omega$. Typical examples of high- and low-resistance nanotube I-V curves are presented in fig. 4 by dashed and dash-doted lines correspondingly. The detailed analysis of obtained I-V curves will be reported elsewhere.

## 4. Discussion and conclusions

The demonstrated procedure of providing local electrical contacts to nano objects has a number of advantages compared to other existing techniques.

Firstly, the object to which the contact is made is free-standing, i.e. it is not supported by a substrate. This is crucial for electrical measurements since it allows the complete exclusion of the effects of current leakage through the substrate and the influence of the substrate on the electronic structure of the sample investigated.

Secondly, conventional techniques of field emission tip preparation and their characterization (by field emission- and field ion microscopy) allow atomically clean tips with a very well defined geometry to be obtained; recently, the preparation of 'nanotips' terminating with protrusions only a few nanometer high and having a single atom sharpness has been made possible[17,18]. Consequently, using these tips for contacting a sample by the described method will provide a very well geometrically and structurally defined atomic-size electrical contact between the tip and the object.

Finally, the process of sample imaging and tip positioning is simple and fast. A rope of any required length and/or diameter can be locally contacted at any required place. In our experiment the measurements of up to ten selected nanotube ropes could be performed with a single field emission tip in approximately one hour.

## Acknowledgements

The authors are grateful to Dr. Z-.C. Dong, Dr. T. Ohgi and Dr. L. D. Ueda-Sarson for fruitful discussions.